# Stern-Gerlach Interferometers with Dual Sensing: Probing Recoherence and Lifecycles of Islands of Coherence

Xing M. Wang[1]

## Abstract

The Branched Hilbert Subspace Interpretation (BHSI) addresses the quantum measurement problem by preserving unitary quantum evolution within a single world. Its central concept is the Island of Coherence (IOC), an operationally isolated, inseparable quantum system mathematically described by a Local Hilbert Space (LHS), which carries no intrinsic spacetime metric and coexists with the spacetime in which the IOC is embedded, a dual structure implicated by the first quantization. This paper advances BHSI on both experimental and conceptual frontiers. Experimentally, we propose a three-stage dual-sensing Stern–Gerlach interferometer (SGI) designed to probe the fuzzy spatiotemporal boundaries associated with IOC transitions. Stage 1 targets uncommitted timing events, manifested as sensor–detector mismatches; Stage 2 investigates conditional recoherence, a signature of local, time-extended branching; and Stage 3 employs controlled electromagnetic phase shifts to discriminate between unitary and retrocausal recoherence mechanisms. Conceptually, we introduce the lifecycle of IOCs, describing how coherent domains emerge, persist, and fragment across scales. We draw structural analogies between fuzzy IOC boundaries and phenomenological bag models in quantum field theory, and between primordial global Hilbert space fragmentation and Hilbert space fragmentation in many-body systems. Altogether, BHSI offers a consistent and experimentally testable approach to resolving the quantum measurement problem.

## 1. Introduction

In our previous articles [1,2], we proposed the Branched Hilbert Subspace Interpretation (BHSI). Within this framework, a measurement is a time-extended sequence of unitary operations—branching, engaging, and disengaging, confined to a *local Hilbert space* (LHS), which is the mathematical description of an operationally isolated island of coherence (IOC), ranging from microscopic systems to astronomical objects. The LHS is characterized by the intrinsic nonlocality of its Hilbert-space structure and coexists with the spacetime in which the IOC is embedded, a dual structure implicated by the first quantization [2]. We proposed experiments [1] using modern full-loop Stern-Gerlach interferometers [3,4] to visualize controlled recoherence processes and branch weights. BHSI preserves unitarity and the Born rule, avoids the wavefunction collapse of Copenhagen Interpretation (CI) [5,6], does not proliferate worlds of

[1] Sherman Visual Lab, Sunnyvale, CA 94085, USA; xmwang@shermanlab.com; ORCID:0000-0001-8673-925X

Many-Worlds Interpretation (MWI) [7,8], and requires no hidden variables of Bohmian mechanics (BM) [9,10].

To probe the ontology and dynamics of measurement, we revisited Einstein's 1927 electron-diffraction thought experiment [11,12] using dual-sensing detector arrays [2]. That work demonstrated how the Born rule can be derived and visualized, and how anomalies, such as mismatched reads ("uncommitted timing events"), can be naturally explained within BHSI's operational sequence and the fuzzy spatiotemporal boundary zone [2].

In this article, we advance BHSI on two interconnected fronts: *experimental* and *conceptual.*

*Experimentally*, we introduce a three-stage investigation using full-loop Stern–Gerlach interferometers (SGIs) [3,4] equipped with *dual sensing* (Sections 2-4), fabricated using modern microfabrication techniques [13-15]. Dual sensors combine non-destructive transparent sensors (TSs) and projective opaque detectors (ODs) to track a single spin-½ particle's path. The stages are designed to test key predictions progressively:

Stage 1 reproduces and verifies anomalies like TS-OD mismatches ("uncommitted timing events"); Stage 2 directly probes conditional recoherence—the potential remerging of branches after a local measurement; Stage 3 adds a controlled electromagnetic phase shift to discriminate between unitary and retrocausal mechanisms for any observed recoherence.

Observation of TS-OD mismatching or recoherence would provide clear empirical signatures to distinguish BHSI from collapse-based or many-worlds interpretations, thereby challenging conventional notions of quantum-classical boundaries and causality [16–19].

*Conceptually*, in Sec. 5, we probe the *lifecycles* of IOCs, given that related quantum systems are not always stable. The IOC's lifecycle, like its boundaries [2], is *operation-dependent*. There are many categories of IOC: from the constantly refreshed IOC for repeating experiments (e.g., Bell tests), to the stable or semi-stable IOC in quantum statistics [20] (e.g., the Einstein-Bose condensation in superconductors and the Hilbert space fragmentation, HSF, in many body systems [21]), to the merging-fragmenting hadronic "bags" (inspired by the phenomenological MIT model [22] and its Lorentz boost [23]) in quantum field theory (QFT) [24], such as the discovery of Higgs bosons at Large Hadron Collider (LHC) [25].

Finally, we propose the possible fragmentation of a primordial global Hilbert space (GHS) in the early universe after electroweak symmetry breaking [26-28], connecting it with the experimental breakthroughs in HSF of many-body systems [21]. We associate the cloudy bag model [29] with the conceptual notion of fuzzy (or semiclassical) spatiotemporal boundaries of IOCs, which is to be probed experimentally using dual-sensing Stern–Gerlach interferometry in the next sections.

## 2. Single Stern-Gerlach Interferometer and Branching Dynamics

In the first stage, we use a single SGI, or the upper half of a full-loop SGI [3]. We insert two compact dual sensors, each comprising a transparent sensor (ST) atop an opaque detector (OD). Hence, the upper SGI forms a closed system (Fig. 1).

## 2.1.Technical Feasibility: TS and OD Separation

The dual sensors can be manufactured as integrated compact measurement units. The core of the manufacturing challenge is to build a single, compact unit where the optical path for the TS is precisely aligned with the particle's trajectory just before it strikes the OD. The minimum distance between the TS's probing region and the OD's active surface could be reduced to sub-millimeter scales with modern microfabrication techniques [13-15]. The limiting factors are the physical size of the components and the need to avoid optical interference.

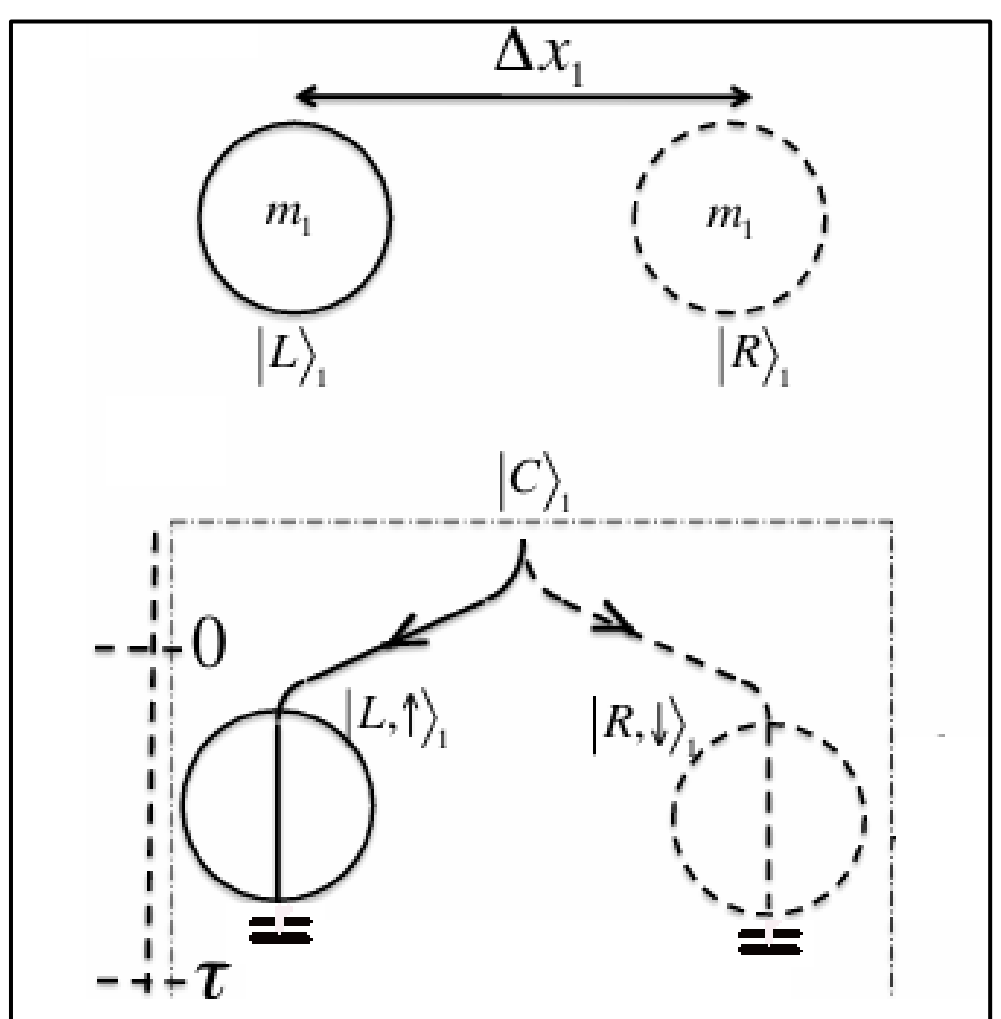

Fig. 1: Single SGI with Two Dual Sensors[2]

The TS is a laser-based optical probe, meaning it is a light beam. Its "footprint" is determined by the size of the laser focus and the collection optics. This can be sub-millimeter. The OD is a physical detector, such as a Microchannel Plate (MCP) or a highly pixelated silicon sensor. These have finite physical thickness. The active surface of an MCP can be placed very close to another object.

The minimum separation distance can be achieved with custom-designed hardware. For example:

- Integrated Detector Assemblies: MCPs can be manufactured in compact stacks with very thin front windows. The laser for the TS could be directed through a small aperture or from the side, allowing the beam to cross the particle's path and strike a photodetector.
- Detector Size: The physical thickness of a single-stage MCP is typically a few millimeters.
- Optical Access: The primary constraint is ensuring the laser beam and the fluorescence photons (if using a fluorescence-based TS) have clear paths without being blocked by the OD's housing. This would require precision engineering of the detector mounts.

[2] The three figures (including this) presented here are modified versions of the original Fig. 1 in Ref. [4].

Given the typical velocities of cold atoms in SGI experiments, a millimeter separation is sufficient for our purpose. The reaction time of the TS is in the nanosecond-to-microsecond range, and the OD can be even faster, making the time-of-flight between them trivial compared to the overall coherence time of the experiment.

The time window $T_W$ (~60 ns) for counting any two successive clicks is set as follows:

$$\tau_{OD}\,(\sim 1\,\text{ns}) < \tau_{TS}\,(\sim 10\,\text{ns}) < T_W\,(\sim 60\,\text{ns}) << 1/f\,(\sim 1\text{ ms}) \tag{1}$$

Such an arrangement introduces measurement timing uncertainty within the fuzzy boundary between the quantum and classical worlds. Due to the setup and measurement timing uncertainties of the dual sensors, along with mostly normal expected outputs, we are likely to observe rare, unexpected, and anomalous results.

**The Initial State**: The ion, located at $|C\rangle_1$, is initialized in the following spin qubit state.

$$|\psi_0\rangle_1 = \cos\theta\,|\uparrow\rangle_1 + e^{i\phi}\sin\theta\,|\downarrow\rangle_1 \tag{2}$$

This can be achieved using rotated magnetic fields or RF pulses [3]. The next step is decoherence: dropping the ion into the vertical SGI, whose gradient magnetic field entangles spin with momentum, forming a superposition of two paths (left and right, see Fig. 1):

$$|\psi_0\rangle_1\,|C\rangle_1 \to |\psi(0)\rangle_1 = \cos\theta\,|\uparrow, L\rangle_1 + e^{i\phi}\sin\theta\,|\downarrow, R\rangle_1 \tag{3}$$

Denote the four readings of the two dual sensors as: [$ST_L$, $ST_R$; $OD_L$, $OD_R$] (entries are 1 = click, 0 = no click). The outputs of ST or OD can be [1,0], [0,1], [0,0], or [1,1]. However, for the TS, it is limited by the mathematical conservation law of total probability (which should sum to 1). In contrast, for the OD, there are physical conservation laws (such as mass, energy, charge, and more), in addition to the total probability.

### 2.2. Normal (Expected) Outcomes

$$[1,0;1,0]\ (|\uparrow\rangle_{ST,L} \to |\uparrow\rangle_{OD,L}) \text{ or } [0,1;0,1]\ (|\downarrow\rangle_{ST,R} \to |\downarrow\rangle_{OD,R}) \tag{4}$$

This is the typical, expected result from a quantum measurement. A hit on a transparent sensor (TS) in one path is immediately followed by a hit on the opaque detector (OD) in the same path. The TS-OD pair on the other path does not produce a click. All three interpretations agree with these normal outcomes, but each offers a different explanation.

**Interpretations:** The output [1,0;1,0] confirms that the left TS acts as a standard measurement device, decohering the particle's wave function and selecting a single, committed path.

- **CI:** The interaction with the left TS causes the system to collapse into the left dual sensor. The collapse is instantaneous and global: the probability of hitting another path vanishes, even if the two sensors are located light-years apart, indicating that collapse is a faster-than-light process.

- **MWI:** No collapse. The interaction with the left TS causes a branching into two worlds.

$$|\psi\rangle_1 |E\rangle_1 \xrightarrow{TS} \cos\theta |\uparrow\rangle_1 |E_\uparrow\rangle_1 + \sin\theta |\downarrow\rangle_1 |E_\downarrow\rangle_1, \quad {}_1\langle E_\uparrow | E_\downarrow\rangle_1 \to 0 \tag{5}$$

In our world, it activates the left dual sensor, while in the other world, it activates the right dual sensor. The global branching is also instantaneous, facing the same issue as CI does.

- **BHSI:** No collapse in a single World. This situation is identical to the Case **5.1**: Aligned Detection (TS #35 → OD #35) in [2]. The interaction with a TS causes the system to branch into a single, locally decohered Hilbert subspace corresponding to that path. If the TS reads 1, the OD registers it immediately thereafter. If the TS reads 0, the decoherent branch is quickly integrated with the environment, thereby preventing the OD from being triggered. For the outcome [1,0; 1,0], we can write:

$$\text{Left}: \quad t_0 : |\uparrow, L\rangle \xrightarrow{\text{ST}} |\uparrow, E_\uparrow\rangle\, |\text{reads} \uparrow\rangle_O \xrightarrow{\text{gap}} |\uparrow, E_\uparrow\rangle \xrightarrow{\text{OD}} |\uparrow, E_\uparrow\rangle\, |\text{reads} \uparrow\rangle_O \to |E'\rangle \tag{6}$$

$$\text{Right}: \quad t_0 : |\downarrow, R\rangle \xrightarrow{\text{ST}} |\downarrow, E_\downarrow\rangle \xrightarrow{\text{gap}} |\downarrow, E_\downarrow\rangle \to |E'\rangle \tag{7}$$

Below the two dual sensors (outside the closed system), there is no longer any probability of finding the ion, given conservation laws (mass, energy, charge, etc.). No spooky action because the two branches belong to a single IOC with an LHS without a spacetime metric.

**The Born Rule:** The frequency with which these outcomes occur, [1,0;1,0] vs. [0,1;0,1], should, over many trials, match the probabilities determined by the initial quantum state. This would provide additional direct verification of the Born rule encoded in the branch weights of BHSI [2].

$$P(\uparrow, L) = P([1,0;1,0]) = \cos^2\theta, \quad P(\downarrow, R) = P([0,1;0,1]) = \sin^2\theta \tag{8}$$

### 2.3. Abnormal (Unexpected) Outcomes

We are interested only in anomalous outcomes that do not violate conservation laws and cannot be explained away by statistical or equipment error. Such anomalies provide a framework for interpreting experimental data and for comparing different interpretations.

**2.3.1. TS-OD Mismatching:** [1,0;0,1] ($|\uparrow\rangle_L \to |\downarrow\rangle_R$) or [0,1;1,0] ($|\downarrow\rangle_R \to |\uparrow\rangle_L$)

A hit on a TS in one path (e.g., $TS_L$) is followed by a hit on the OD in the *other* path ($OD_R$).

**Interpretation:** Although these abnormal outcomes are rare, they do not violate any conservation laws. They may occur because the TS reaction time is less or comparable to the particle transition time between TS and OD, plus the OD reaction time. The closer the TS and OD are, the shorter the OD reaction time, and the higher the chance of it happening.

- **CI:** It is difficult to explain: Instant collapse fixes subsequent outcomes. If truly observed, the standard CI assumptions are violated.

- **MWI:** This outcome is also difficult to account for, as the TS measurement should have irreversibly split the universe. The final OD detection should then be aligned with the TS detection in that specific world. If swapping truly happens, the standard assumptions of MWI are violated.

- **BHSI ("uncommitted timing events"):** It can be explained naturally. For example, the outcome [1,0; 0,1] can be written as:

$$\begin{aligned} &\text{Left}: && t_0 :|\uparrow, L\rangle \xrightarrow{\text{ST}} |\uparrow, E_\uparrow\rangle\,|\,\text{reads}\uparrow\rangle_O\ t_2 : \xrightarrow{\text{gap}} |\,E'\rangle \\ &\text{Right}: && t_0 :|\downarrow, R\rangle \xrightarrow{\text{ST}} |\downarrow, E_\downarrow\rangle \xrightarrow{\text{gap}} |\downarrow, E_\downarrow\rangle \xrightarrow{\text{OD}} t_1 :|\downarrow, E_\downarrow\rangle\,|\,\text{reads}\downarrow\rangle_O \rightarrow |\,E'\rangle \end{aligned} \qquad (t_0 < t_1 < t_2) \qquad (9)$$

The TS-OD mismatch, an anomaly in which transparent sensors and opaque detectors disagree, mirrors the misalignment case (e.g., TS #35 → OD #45) in Section **5.2** of [2]. Although anomalous, it does not violate any conservation laws. Both cases indicate that quantum systems can exhibit temporally local branch resolution, implying a fuzzy boundary between the quantum and classical worlds in dual-sensing, where subsystem measurements remain uncertain until a subsequent interaction enforces global consistency.

Although the temporal anomalies resemble Wheeler's delayed-choice scenarios [16-19], within BHSI they are interpreted *not as retrocausal* effects but as uncommitted timing events in a 'fuzzy' zone, reflecting the finite duration of the local branching process.

**2.3.2. No TS reading:** [0,0; 0,1] or [0,0,1,0].

Neither TS has fired. The dual sensors behave simply like ODs, implying local and timely dynamic branching, as in **5.3** of [2]: missing inner detection. But whether global probability conservation could be temporarily violated in the fuzzy zone is still an open question

### 2.3.3. Anomalous Readings Violating Conservation Laws

Similar to what we did in [2], in this and the next two stages, scenarios involving apparent violations of energy or particle conservation (e.g., duplicate detections on TS or OD only) are excluded from consideration, because such events would indicate experimental artifacts or a breakdown of quantum theory itself rather than differences between interpretations

## 3. Single Full-Loop Stern-Gerlach Interferometer and Recoherence

According to BHSI, after measurement, the branches evolve unitarily and independently. Their recoherence is possible if they remerge before being permanently relocated in the local environment, provided that no conservation laws are violated (e.g., energy, mass, charge). To explore the unique feature of BHSI, we extended the experimental setup described above by adding the lower half of the full SGI. We insert two TSs between the upper and lower SGI, and use an OD to detect the final state at the bottom center of the lower SGI (Fig. 2).

To increase the likelihood of "recoherence," the crucial condition is that the acceleration or deceleration time, Δt, is shorter than or comparable to the TS reaction time (1 μs ~ 1 ns). The data used in the SGI gravity test are [4]: $\Delta x \sim 10$ μm, $\partial B \sim$ 10^6 T, $m \sim 10^{-14}$ kg, and $\Delta t \sim 0.1$s. Keeping $\Delta x$ and $\partial B$ unchanged, we can estimate $\Delta t$ as follows:

$$\Delta x \sim (F_{\partial B} / m)\Delta t^2 / 2 \rightarrow \Delta t \propto \sqrt{m} \rightarrow \Delta t_e \propto \Delta t\sqrt{m_e / m} \qquad (10)$$

Therefore, we choose electrons as our charged particle to reduce $\Delta t$. In alignment with existing high-gradient SGI experiments, we maintain a magnetic field gradient of $\partial B \sim 10^6$ T/m. For electrons, this yields a characteristic acceleration time of $\Delta t_e \sim 1$ ns, which is comparable to the response time of state-of-the-art transparent sensors.

**The Initial and the controlled decoherent state**: The electron, located at $|C\rangle_1$, is initialized in the following spin qubit state.

$$|\psi_0\rangle_1 = \cos\theta\,|\uparrow\rangle_1 + e^{i\varphi}\sin\theta\,|\downarrow\rangle_1 = (|\uparrow\rangle_1 + |\downarrow\rangle_1)/\sqrt{2} \quad (\theta = \pi/4, \varphi = 0) \qquad (11)$$

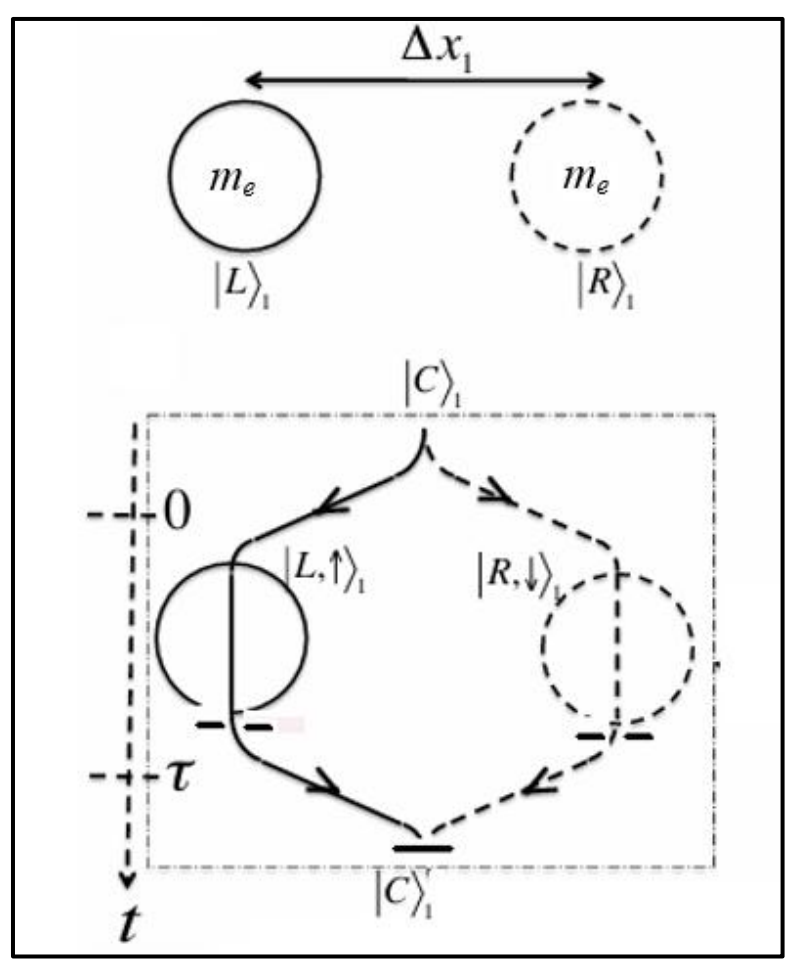

Fig.2: Single Full-Loop SGI with Two TSs and One OD

The next step is to change it to a locally controlled *decoherent state*: drop the electron vertically into the upper SGI, whose gradient magnetic field entangles spin with momentum, forming a superposition of two paths (left and right) [1, 4]:

$$|\psi(0)\rangle_1 = \cos\theta\,|\uparrow, L\rangle_1 + e^{i\varphi}\sin\theta\,|\downarrow .R\rangle_1 = (|\uparrow, L\rangle_1 + |\downarrow, R\rangle_1)/\sqrt{2} \quad (\theta = \pi/4, \varphi = 0) \qquad (12)$$

Without inserting the two TSs between the upper and lower SGI, the electron moves down to the lower SGI, and its two branches remerge together by the opposite magnetic gradient. If we measure the electron's state with the OD at the bottom center, we should find that the final state is identical to its initial state, Eq. (11) [1, 4]:

$$|\psi_f\rangle_1 = \cos\theta|\uparrow\rangle_1 + e^{i\varphi}\sin\theta|\downarrow\rangle_1 = (|\uparrow\rangle_1 + |\downarrow\rangle_1)/\sqrt{2} \quad (\theta = \pi/4, \varphi = 0) \tag{13}$$

This is the Controlled Decoherence-Recoherence Process (CDRP), described in Section **5.3.1**, Eq. (52) of Ref. [1], with the specialized values of $\theta$ and $\varphi$.

We now insert the two TSs between the upper and lower SGI (Fig. 2). The two TSs and the OD will yield different readings. Let us denote the output as: [$TS_L$, $TS_R$; $\theta'$]. For example, [1,0; 0] means the left TS fires and OD reads $\theta' = 0$. The final state can be written as:

$$[1,0;0]:\ |\psi_f\rangle_1 = \cos\theta'|\uparrow\rangle_1 + e^{i\varphi'}\sin\theta'|\downarrow\rangle_1 = |\uparrow\rangle_1 \quad (\theta' = 0, \varphi' = 0) \tag{14}$$

The normal (expected) and anomalous (unexpected) outputs are listed as follows.

### 3.1. The Normal Outputs:

[1,0; 0] → |↑⟩, or [0,1; π/2] → |↓⟩,

**Interpretations:** all three are expected, although each offers a different explanation.

**CI**: The fire of $TS_L$ ($TS_R$) collapses the wave function to a fixed spin state |↑⟩ (|↓⟩), the other path (with opposite spin state) disappears instantly, then the OD confirms the fixed spin state by reading $\theta' = 0$ ($\pi/2$).

**MWI**: The fire of $TS_L$ splits the current world into two global branches (worlds).

$$|\psi\rangle_1|E\rangle_1 \xrightarrow{TS} (|\uparrow\rangle_1|E_\uparrow\rangle_1 + |\downarrow\rangle_1|E_\downarrow\rangle_1)/\sqrt{2}, \quad {}_1\langle E_\uparrow|E_\downarrow\rangle_1 \to 0 \tag{15}$$

In our world, the left TS reads one (a fixed spin state |↑⟩), the right TS reads zero simultaneously, and the OD confirms the choice by reading $\theta' = 0$. At the same time, in the other world (which we cannot communicate with), it is the $TS_R$ that reads one, choosing spin state |↓⟩, while $TS_L$ reads zero, and its OD confirms the choice by reading $\theta' = \pi/2$.

**BHSI**: No collapse, single World. The $TS_L$ fire splits the local Hilbert space into two branches. In the left branch, the spin state is fixed at |↑⟩, and the OD confirms this choice by reading $\theta = 0$. Meanwhile, in the other branch, although its $TS_R$ reads zero (without engaging with the sensor), it does not disappear immediately; it continues to evolve as an independent, decoherent branch for a short period before finally integrating into its environment without triggering the OD.

$$|\uparrow, L\rangle/\sqrt{2} \xrightarrow{TS_L} |\uparrow_B, E_\uparrow\rangle|\text{reads } \uparrow\rangle/\sqrt{2}_L \xrightarrow{OD} |\uparrow\rangle; \quad |\downarrow, R\rangle/\sqrt{2} \xrightarrow{TS_R} |\downarrow_B, E_\downarrow\rangle/\sqrt{2} \to |E'\rangle \tag{16}$$

### 3.2. Abnormal (Unexpected) Outcomes

There are potentially different abnormal outcomes. Some of them are prohibited by global conservation laws (e.g., mass, energy, charge, total probability), which we will not discuss here.

**3.2.1. Anomalous OD Reading (Remerged Superposition):** [1,0; $\pi/4$] or [0,1; $\pi/4$].

The left or right TS fires, selecting a spin-up or spin-down state, whereas the OD detects a superposition of both states.

**Interpretations:** This anomaly does not violate any conservation laws and can clearly distinguish BHSI from CI and MWI.

**CI**: The fire of $TS_L$ ($TS_R$) collapses the wave function to a fixed spin state $|\uparrow\rangle$ ($|\downarrow\rangle$), the other path (with the opposite spin state) disappears instantly, so there is no chance that OD finds a superposition of the two spin states. It violates CI's fundamental assumption.

**MWI**: The fire of $TS_L$ splits the current world into two global branches (worlds) as in Eq. (15). Each spin state appears in its own world. There is no chance that the OD finds a superposition of the two spin states. This violates the fundamental assumption of MWI.

**BHSI** (**"Recoherence"**): The fire of $TS_L$ splits the local Hilbert space into two local branches. In the left branch, the TS detects the fixed spin state $|\uparrow\rangle$. Meanwhile, the right branch passes through the TS without detection and continues to evolve as an independent branch. Then the OD finds a superposition of the two spin states by reading $\theta' = \pi/4$. It means that the two branches are merged back by the magnetic gradient of the lower SGI – a signal of *recoherence*, consistent with the fundamental assumptions of BHSI.

$$|\uparrow, L\rangle/\sqrt{2} \xrightarrow{TS_L} |\uparrow, E_\uparrow\rangle \,|\, \text{reads } \uparrow\rangle/\sqrt{2}_L \rightarrow \text{OD}; \quad |\downarrow, R\rangle/\sqrt{2} \xrightarrow{TS_R} |\downarrow, E_\downarrow\rangle/\sqrt{2} \rightarrow \text{OD} \tag{17}$$

**3.2.2. TS-OD Mismatching**: [1,0; $\pi/2$] or [0,1; 0]

"Uncommitted timing events." Similar to the case in Section **2.3.1**: [1,0; 0,1] or [0,1; 1,0].

## 4. Two Full-Loop Stern-Gerlach Interferometers, Recoherence, and Retrocausality

Wheeler's delayed-choice thought experiment [16-19] highlights the tension between quantum measurement and temporal causality: If a photon's wave/particle behavior is determined after it has traversed the interferometer, does the choice "retrocausally" influence its past state? Two interpretations arise:

- *No Retrocausality*: The outcome is determined by the experimental configuration at the time of detection (consistent with Copenhagen's "measurement-induced collapse").
- *Retrocausality*: The future measurement choice affects past dynamics (e.g., time-symmetric or transactional interpretations).

If recoherence is detected in the single full-SGI experiment, our two-SGIs setup can be extended to detect both recoherence and the EM phase shifts ($\Delta\Phi$). By adding another full-loop SGI on the right with a falling charged particle and keeping the setup of the left full-loop SGI as in the second stage (Fig. 3), we then use the opaque detector (OD) to measure the final state and check whether it is a recohered state (a superposition) and whether it has the electromagnetic (EM) phase shift, which was estimated as in Eq. (56) of [1]:

$$\Delta\Phi(\tau)=\frac{q_1q_2\tau}{\hbar}\left(\frac{1}{d-\Delta x_1/2}-\frac{1}{d+\Delta x_1/2}\right)\sim\frac{q_1q_2\tau\Delta x_1}{\hbar[d^2-(\Delta x_1)^2/4]}\sim\frac{q_1q_2\tau\Delta x_1}{\hbar d^2}\sim 0.2\text{ rad} \tag{18}$$

Here, we assume $q_1 \sim q_2 \sim -3e$, $d \sim 100$ μm, $\Delta x \sim 10$ μm, and $\tau \sim 100$ ms. In alignment with existing high-gradient SGI experiments [4], we assume a magnetic field gradient on the order of $\partial B \sim 10^6$ T/m. For electrons, this yields a characteristic acceleration time of $\Delta t \sim 1$ ns, which is comparable to the response time of state-of-the-art transparent sensors. If we use electrons on both SGIs, the resulting electrodynamic (EM) phase shift will be reduced to $\Delta\Phi \sim 0.02$ rad, which remains within current interferometric measurement sensitivity.

To enhance the electromagnetic phase shift in the dual-SGI setup, we propose replacing the right-arm electron with a heavy spin-0 ion with a mass $\sim 10^4$ times that of an electron and a charge of $\sim -5e$. This ion is not subject to "splitting" and falls freely alongside the left-arm electron. Its motion generates a relatively static electric field that interacts with the decoherent branches of the left SGI. As a result, the EM phase shift experienced by the left electron increases to $\Delta\Phi \sim 0.1$ radians, compared to $\Delta\Phi \sim 0.02$ radians when both SGIs employ electrons.

$$\Delta\Phi(\tau)\sim\frac{q_1q_2\tau\Delta x_1}{\hbar[d^2-(\Delta x_1)^2/4]}\sim\frac{5e^2\tau\Delta x_1}{\hbar d^2}\sim 0.1\text{ rad} \tag{19}$$

This increase significantly improves the detectability of potential recoherence anomalies arising from EM phase shifts. Importantly, no measurements are made on the heavy ion, which acts solely as a moving source of the electric field, influencing the left-side quantum system (Fig. 3).

The initial state is set to Eq. (11) for the single-SGI setup. It splits into two branches due to the magnetic gradient of the left upper SGI, then falls into the space between the upper and lower SGI, experiencing an electric force from the synchronously falling heavy ion on the right full-loop SGI. Without inserting the two TSs, the two branches merge due to the magnetic gradient, forming a superposition with an EM phase shift:

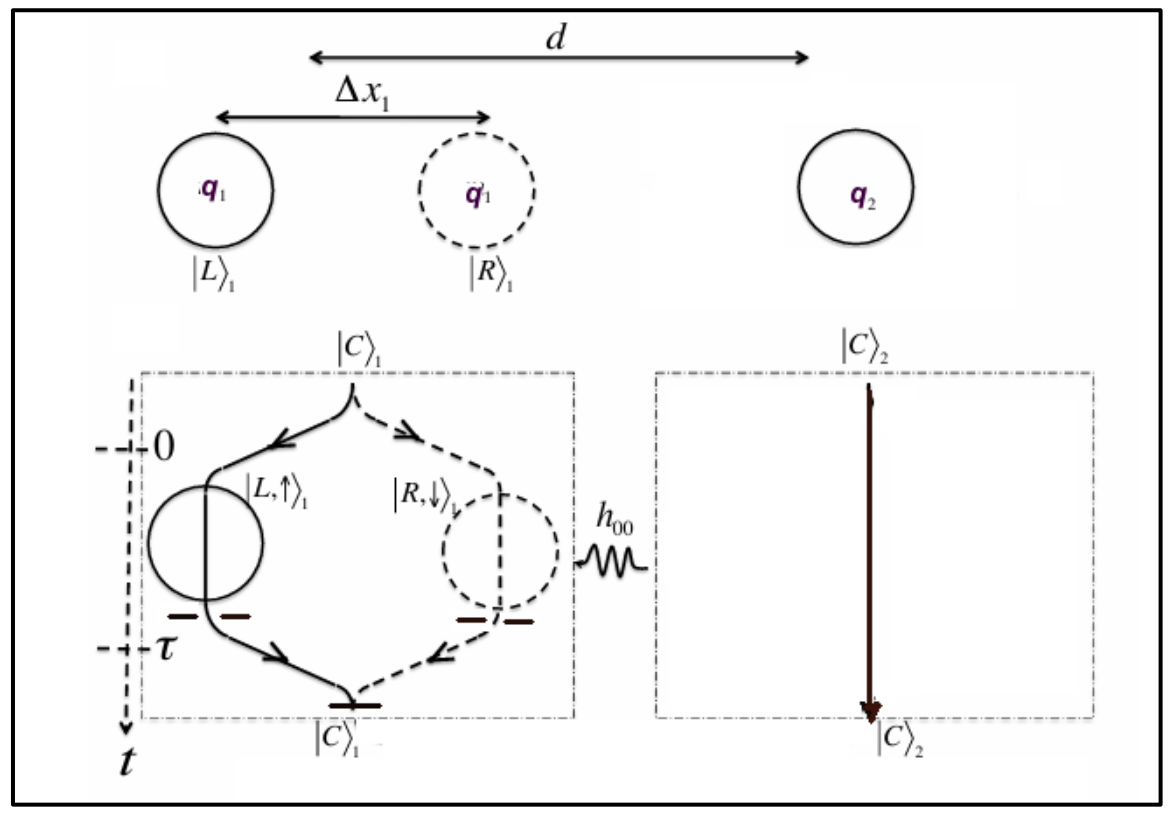

Fig.3: Two Full-Loop SGIs with Two TSs and One OD on the Left SGI

$$|\psi_0\rangle_1=(|\uparrow\rangle_1+\beta|\downarrow_1\rangle)/\sqrt{2}\rightarrow U(t)(|\uparrow,L\rangle_1+|\downarrow,R\rangle_1)/\sqrt{2}\overset{t=\tau}{\rightarrow}e^{i\Phi_L(\tau)}[|\uparrow\rangle_1+e^{i\Delta\Phi(\tau)}|\downarrow\rangle_1]/\sqrt{2} \tag{20}$$

Denoting the outcome by [$TS_L$, $TS_R$; θ, φ]. After inserting the two TSs at the entrance of the lower SGI, we will observe the following outcomes from the two TSs and the OD.

**4.1. Normal**: [1,0; 0, 0] or [0,1; $\pi$/2,0].
Similar to the normal case **3.1**.

**4.2. Anomalous**:

There are two interesting anomalies.

**4.2.1**. Anomalous OD Reading with Merging (Recoherence)**:**
Similar to case **3.2.1.**
Without the phase shift: [1,0; $\pi$/4,0] or [0,1; $\pi$/4,0]
With the phase shift: [1,0; $\pi$/4, ΔΦ] or [0,1; $\pi$/4, ΔΦ]

**4.2.2.** TS-OD Mismatching: [1,0; $\pi$/2, 0] or [0,1; 0, 0].
Similar to the cases in **2.3.1** and **3.2.4**: "Uncommitted timing events."

The most interesting case is 4.2.1, the anomalous OD reading with remerged superposition-recoherence. It does not violate any conservation laws, yet clearly distinguishes CI and MWI from BHSI and validates the principle of non-retrocausality in quantum mechanics. If the OD observes no phase shift ΔΦ, it implies that the TS's measurement erased the independent unitary evolution of the two branches, a signature consistent with retrocausality (where the future measurement choice disrupts past coherence). Conversely, if the OD observes ΔΦ, it demonstrates that the phase shift is fully determined by local unitary dynamics, independent of the TS's actions. This would falsify retrocausal interpretations, affirming that the system's evolution is governed by the pre-measurement history of quantum components [16-19].

## 5. The Lifecycles of Local Hilbert Spaces

We have introduced the islands of coherence (IOC) ontology to describe local Hilbert spaces (LHSs) [2]. Because an observed quantum system must be sufficiently isolated from environmental decoherence to sustain coherent unitary evolution, effective quantum boundaries arise that separate the system from surrounding classical systems and from other quantum systems with which it does not form a single inseparable whole. IOCs are described by their associated LHSs, coexisting with the background spacetime [2]. These quantum systems are not necessarily dynamically stable or at rest. Consequently, IOCs may emerge, co-move with associated quantum systems, persist, and dissolve or fragment over time. The specific form of this lifecycle is not intrinsic to the system; rather, it is defined by the observational context.

In this section, we discuss the lifecycles of IOCs across different physical contexts. One must recognize that both the lifecycles and boundaries of IOCs are operation-dependent. For example, observing the optical spectrum of a Uranium-235 atom may involve the IOC of the atomic electrons; probing its nuclear structure would invoke the IOC of the nucleons, which may split into environmentally decoherent IOCs after a neutron-induced nuclear fission; and investigating the quark-gluon structure inside a proton (as in electron-proton scattering) may invoke the IOC of the proton, which in turn can fragment by high-energy collision with another proton [25]. In each case, the relevant IOC is defined by the rate of information leakage to the environment or by the resolution of the measurement operation [2], rather than by a fixed ontological partition.

### 5.1. IOCs of Repeating Quantum Measurements

Because a quantum measurement can yield different outcomes, with probabilities governed by the Born rule, experiments must typically be repeated hundreds or thousands of times under nominally identical conditions. The electron diffraction experiment discussed in Ref. [2] provides a representative example. Inside the sealed hemispherical detector, an IOC exists that begins with the arrival of the first electron and ends with the detection of the last.

Operationally, one may also regard the IOC as effectively refreshed by each incoming electron during the experimental run, without any observable distinction between these descriptions. This illustrates that branching, engagement, and disengagement are processes local to this transient IOC, without affecting the apparatus outside it.

The Stern–Gerlach interferometry (SGI) experiments analyzed in this article provide another example, as do Bell-type tests involving entangled photon pairs. In all such cases, repeated trials probe statistical branch weights for a single experimental configuration, rather than requiring a distinct global Hilbert space for each outcome.

### 5.2. IOCs of Identical Particles and Quantum Statistics

Many of the IOC examples (Sec. 6.4 of [2]) involve systems composed of multiple identical particles, such as Cooper pairs in superconducting tunneling or neutrons in a neutron star. These systems consist of either fermions or bosons and form inseparable quantum islands. No experiment can localize an individual particle or trace its trajectory without destroying the system's coherence.

The indistinguishability of identical particles plays a central role in quantum statistics, giving rise to phenomena such as Fermi degeneracy and Bose–Einstein condensation [20]. Within BHSI, these statistical effects arise naturally from the structure of a single, inseparable IOC rather than from correlations among distinguishable particles. Such systems, together with their associated LHSs, remain stable or semi-stable as long as the physical conditions supporting coherence are maintained. Otherwise, the IOC may transform. For example, as a white dwarf's mass increases, it may collapse into a neutron star, which in turn may collapse into a black hole.

The lifetimes of many-body IOCs may be fundamentally linked to the phenomenon of Hilbert space fragmentation (HSF), which has recently been observed in two-dimensional systems [21]. In such systems, kinetic constraints give rise to dynamically decoupled subsectors, “islands” within the Hilbert space, that evade thermalization and remain non-ergodic. This empirical observation of HSF provides strong support for the BHSI view that the total Hilbert space of a quantum many-body system can fragment into separable LHSs, determined by the system's internal dynamics and symmetries.

### 5.3. IOCs in Quantum Field Theory

In the momentum representation (Sec. 3.1-3.6, [24]), the basis states are momentum eigenvectors $|p\rangle$, and different momentum eigenstates are mutually orthogonal. In the position ($x$) representation, a momentum eigenstate yields a constant probability distribution, reflecting its complete spatial delocalization (a manifestation of the Heisenberg uncertainty principle):

$$\langle x \mid p\rangle = e^{i x \cdot p/\hbar} / \sqrt{2\pi\hbar}, \quad |\langle x \mid p\rangle|^2 = 1/2\pi\hbar \tag{21}$$

In quantum field theory (QFT), interacting particles are described by occupation numbers in 4-momentum modes of the Fock space. Their interactions are represented by Feynman diagrams (Sec 6.1-6.5, [24]), which depict incoming and outgoing particles with their 4-momenta without localized trajectories; the observable quantity is the differential cross section dσ/dΩ, computed directly from asymptotic 4-momentum states. Because the formalism refers only to initial and final states and global conservation laws, the interaction is naturally described by a single IOC associated with the scattering process. Therefore, nonlocality is also intrinsic to the LHS of QFT, arises from its lack of a spacetime metric, and its coexistence with the background spacetime [2].

In high-energy proton–proton collisions, such as those conducted at the Large Hadron Collider (LHC) [25], hundreds of particles may emerge from a localized interaction region. From the perspective of BHSI, this region may be interpreted as a transient IOC, a "black box" into which two protons repeatedly enter, interact, and produce a large number of outgoing particles that subsequently decohere and are detected by macroscopic sensors. Heuristically, each incoming proton may be viewed as a confined system of quarks and gluons (partons), or a "bag," in the spirit of the phenomenological bag model [22, 23]. During a high-energy collision, these confined systems effectively merge into a unified composite IOC, which then rapidly fragments into numerous bags (hadrons) and other particles [25], moving away in various directions (a local branching in BHSI). Upon interaction with macroscopic detectors, these entities undergo decoherence and transition to the classical regime.

In this way, BHSI provides a relativistic, unitary, and single-world account of QFT processes that remains fully compatible with the standard formalism and its experimental verification.

### 5.4. The Global Hilbert Space and Its Fragmentation in the Early Universe

After the Planck epoch ($\sim 10^{-43}$ s) and the subsequent inflation epoch ($\sim 10^{-32}$ s), spacetime dynamics are well described by classical general relativity. In the electroweak symmetry-breaking epoch ($\sim 10^{-12}$ s), the Standard Model gauge symmetry underwent spontaneous symmetry breaking via the Higgs mechanism, giving masses to elementary particles and qualitatively altering the interaction structure. Prior to this transition, particles were effectively massless and propagated at the speed of light within a highly symmetric interaction regime [24–27]

During this early, unified phase, the Universe can be reasonably modeled as a single quantum system represented by a universal Island of Coherence (IOC) with a global Hilbert space (GHS), which coexists with an effectively classical spacetime. When the Higgs field acquired a uniform, nonzero vacuum expectation value ($\sim 10^{-12}$ s), spontaneous symmetry breaking occurred throughout the whole IOC. Particles acquired mass, characteristic propagation scales changed, and the structure of interactions diversified. Subsequently, during the quark epoch (~$10^{-12}$–$10^{-6}$ s), the Universe consisted of a hot quark–gluon plasma in which bound states could not yet form (see Sec. 3.5 of [27] and its associated data table).

From the perspective of BHSI, this epoch represents a "semiclassical time zone" characterized by the progressive emergence of multiple Local Hilbert Spaces (LHSs). This resulted from the effective Hilbert Space Fragmentation (HSF) of the primordial GHS as the Universe expanded, cooled, and its interaction structure diversified. This picture is conceptually consistent with recent research on Hilbert space fragmentation in many-body systems, such as 2D lattice systems [21]. The regions separating these coherent quantum islands became increasingly well-described by effectively classical spacetime degrees of freedom. In this way, the present-day structure of separated IOCs—existing within an effectively classical "ocean" of metric spacetime—arises naturally from the initially unified primordial IOC.

**Note:** This discussion extends BHSI within the established Standard Cosmological Model timeline (Friedmann's equation) [28]. The primordial IOC and its subsequent fragmentation into IOCs are taken to coexist consistently with classical spacetime, which evolves under general relativity in the quark epoch. No specific theory of quantum gravity is assumed or required.

### 5.5. Fuzzy Spatiotemporal Boundary of LHSs

In all the above IOC lifecycles, there is an outer boundary, beyond which degrees of freedom are considered decohered and separable from the quantum system, and an inner boundary, within which all components evolve as a single unitary whole. These two boundaries are generally not coincident; there is no sharp, instantaneous cut between the classical world and a quantum island of coherence, as we have discussed in Subsection 6.1.2 of [2]. The absence of a sharp system–environment boundary is also familiar in quantum field theory, for example, in the cloudy bag model [29], where mesonic degrees of freedom extend beyond the nominal confinement region, effectively softening the bag boundary.

In the dual-sensing SGI examples, the fuzzy spatiotemporal zone is located between the TS sensor's entry and the OD's bottom. Our setup is explicitly designed to probe the dynamical events, such as uncommitted timing events or conditional recoherence, that occur within this semiclassical spatiotemporal zone.

## 6. Summary and Discussion

The central concept of Branched Hilbert Subspace Interpretation (BHSI) is the Island of Coherence (IOC), an operationally isolated, inseparable quantum system, mathematically described by a Local Hilbert Space (LHS) that has no intrinsic spacetime metric and coexists with the spacetime in which the IOC is embedded, a dual structure implicit in first quantization. The IOC has an operation-dependent semiclassical spatiotemporal boundary zone, within which a sequence of unitary operations occurs [1,2].

In this article, we have advanced the Branched Hilbert Subspace BHSI on two fronts: an experimental program to test its dynamical predictions and a conceptual expansion grounding its ontology across modern physics.

*Experimental program*. Using full-loop Stern-Gerlach interferometers with dual sensing—combining non-destructive transparent sensors and projective opaque detectors—we can directly probe BHSI's time-extended measurement process. The three-stage architecture targets key signatures: Stage 1 seeks "uncommitted timing events" (TS-OD mismatches), revealing the fuzzy spatiotemporal boundary of local branching; Stage 2 probes conditional recoherence, where locally decohered branches remerge—a prediction distinguishing BHSI from irreversible collapse or global splitting; Stage 3 adds a controlled phase shift to discriminate unitary from retrocausal recoherence mechanisms. Observing these effects would provide empirical criteria distinguishing BHSI from Copenhagen and Many Worlds interpretations.

*Conceptual expansion*. We introduced the *lifecycle* of islands of coherence (IOCs): IOCs may not be static but undergo emergence, persistence, and fragmentation, governed by operational context. This lifecycle perspective unifies quantum phenomena from repeated laboratory measurements to the early evolution of the Universe. The IOC's fuzzy boundary and lifecycle are operation-dependent; drawing on precedents like the cloudy bag model [29], we argue that the quantum-classical transition is an extended, observable process.

*Cosmological extension*. BHSI offers a consistent account of the early Universe: an initially unified Global Hilbert Space (GHS) progressively fragmented during the interval from approximately $10^{-12}$–$10^{-6}$ seconds after the Big Bang, following Higgs-mediated mass acquisition [27]. This picture aligns with recent observations of Hilbert space fragmentation in 2D many-body systems [21], suggesting that today's structure of localized quantum "islands" embedded in classical spacetime is the stable outcome of this dynamical evolution.

*Operationally layered structure*. BHSI reveals a multilayered structure of coherence: the classical world emerges as the macroscopic limit of nested unitary domains. Increasing measurement energy does not merely access smaller spatial scales; it engages progressively deeper layers of unitary coherence—from atomic to nuclear to hadronic IOCs—each defined by its operational context. In this view, IOCs, whether natural or artificial, are objective physical structures independent of human observation, while the phenomena accessible from them depend on the operational parameters of measurement.

Overall, BHSI replaces wavefunction collapse, global world-splitting, and hidden nonlocal dynamics with a continuous, locally unitary, and experimentally testable account of quantum measurement, and remains consistent from laboratory experiments to the thermal history of the Universe.

## Abbreviations

| | |
|---|---|
| BHSI | Branched Hilbert Subspace Interpretation |
| CI | Copenhagen Interpretation |
| GHS | Global Hilbert Space |
| HSF | Hilbert Space Fragmentation |
| IOC | Island of Coherence |
| LHC | Large Hadron Collider |
| LHS | Local Hilbert Space |
| MWI | Many-Worlds Interpretation |
| OD | Opaque Detector |
| QFT | Quantum Field Theory |
| SGI | Stern-Gerlach Interferometer |
| TS | Transparent Sensor |